\documentclass[conference]{IEEEtran}
\DeclareUnicodeCharacter{211D}{\,}
\usepackage[T1]{fontenc}    
\usepackage{hyperref}       
\usepackage{url}            
\usepackage{booktabs}       
\usepackage{amsfonts}       
\usepackage{nicefrac}       
\usepackage{microtype}      
\usepackage{lipsum}

\usepackage{filecontents}
\usepackage{comment}
\usepackage{enumerate}
\usepackage{amsfonts}
\usepackage{amssymb}
\usepackage{verbatim}
\usepackage{setspace}
\usepackage{mathrsfs}
\usepackage{tikz}
\usetikzlibrary{arrows,shapes}
\usepackage{graphicx}

\usepackage{subcaption}

\usepackage{etoolbox}

\usepackage{listings}
\lstdefinelanguage{encryption}
 {morekeywords=X, 
  otherkeywords={1,2,3,4,5,6,7,8,9,0,a,b,c,d,e,f},
  }

\usepackage{todonotes}
\usepackage{bm} 
\usepackage{float}
\usepackage{graphicx}
\usepackage{tikz}
\usetikzlibrary{shapes.geometric}
\usetikzlibrary{arrows}
\usetikzlibrary{arrows.meta,calc,decorations.markings,math,arrows.meta}

\usepackage{amsmath}
\usepackage{mathdots}
\usepackage{diagbox}
\usepackage{amsfonts}
\usepackage{amssymb}

\ifCLASSINFOpdf
\else
\fi
\DeclareRobustCommand*{\IEEEauthorrefmark}[1]{%
\raisebox{0pt}[0pt][0pt]{\textsuperscript{\footnotesize\ensuremath{#1}}}}

\begin{document}
%
\title{Unveiling Decentralization: A Comprehensive Review of Technologies, Comparison, Challenges  \\ in Bitcoin, Ethereum, and Solana Blockchain}

\author{
\IEEEauthorblockN{ 
Han Song\IEEEauthorrefmark{1},
Yihao Wei\IEEEauthorrefmark{2},
Zhongche Qu\IEEEauthorrefmark{3},
Weihan Wang\IEEEauthorrefmark{4}}
\IEEEauthorblockA{\IEEEauthorrefmark{1} University of Southern California, Los Angeles, US}
\IEEEauthorblockA{\IEEEauthorrefmark{2} University of Illinois Urbana-Champaign, Champaign, US}
\IEEEauthorblockA{\IEEEauthorrefmark{3} Columbia University, New York, US}
\IEEEauthorblockA{\IEEEauthorrefmark{4} Vanderbilt University, Nashville, US}
\IEEEauthorblockA{hsong427@usc.edu\IEEEauthorrefmark{1}, yihaow4@illinois.edu\IEEEauthorrefmark{2}, zq2172@columbia.edu\IEEEauthorrefmark{3}, weihan.wang@vanderbilt.edu\IEEEauthorrefmark{4}}
}



\maketitle
\begin{abstract}
Bitcoin stands as a groundbreaking development in decentralized exchange throughout human history, enabling transactions without the need for intermediaries. By leveraging cryptographic proof mechanisms, Bitcoin eliminates the reliance on third-party financial institutions. Ethereum, ranking as the second-largest cryptocurrency by market capitalization, builds upon Bitcoin's groundwork by introducing smart contracts and decentralized applications. Ethereum strives to surpass the limitations of Bitcoin's scripting language, achieving full Turing-completeness for executing intricate computational tasks. Solana introduces a novel architecture for high-performance blockchain, employing timestamps to validate decentralized transactions and significantly boosting block creation throughput. Through a comprehensive examination of these blockchain technologies, their distinctions, and the associated challenges, this paper aims to offer valuable insights and comparative analysis for both researchers and practitioners. 
\end{abstract}


%
\IEEEpeerreviewmaketitle

\section{Introduction}
Mainstream frameworks in use today continue to lean heavily on centralization. For instance, in robotics, we find applications~\cite{song2024bundledslam, xanthidis2022towards, xanthidis2021towards, chen2019visual}  and in AI4Science~\cite{lin2023comprehensive, Lin_2023_ICCV}, as well as internet applications built on centralized Web2 architectures.
Cryptocurrency, originating with the advent of Bitcoin in 2008 \cite{nakamoto2008bitcoin}, represents a modern manifestation of humanity's longstanding pursuit of decentralized exchange. The concept of decentralization itself boasts a historical trajectory spanning over two centuries \cite{cao2022decentralized}, and it has been intricately interwoven into various academic disciplines, including economics, political science, and computer science. The proliferation of decentralized concepts gained momentum with the emergence of Bitcoin in 2009. However, despite the surge of interest and capital inflow into the cryptocurrency market, a significant proportion of participants exhibit a limited understanding of the fundamental principles of decentralization and blockchain technology.

This paper aims to thoroughly analyze the workings of decentralized exchanges, explaining the algorithms used to overcome challenges commonly managed by central authorities. Adopting a chronological approach, we examine the evolutionary progression of blockchain technology by scrutinizing the architectures and protocols of Bitcoin, Ethereum, and Solana. Furthermore, this study delves into an in-depth exploration of the inherent limitations of these platforms relative to their respective offerings.

\section{Bitcoin - Origin of cryptocurrency}
Introduced in 2008, Bitcoin stands as a seminal milestone in the evolution of decentralized exchange within human history. It adeptly facilitates transactions without the necessity of a central trust party, which traditionally acts as an intermediary connecting diverse clients. This electronic payment system circumvents the involvement of third-party financial institutions through the utilization of cryptographic proof mechanisms.

\subsection{Double spending problem}
The double spending problem constitutes a primary challenge in digital currency transactions. Unlike physical currency, such as coins and bills, which are minted by authorized financial institutions and possess inherent scarcity, digital assets are susceptible to duplication. This ease of replication can result in undesirable scenarios wherein a sender may send identical digital coins to multiple recipients, leaving the recipients unaware of potential simultaneous transactions to different addresses. In the absence of a central authority, a payer initiating a transaction must publicly broadcast this action to all participants. Subsequently, a consensus mechanism among the majority is required to validate the transaction's completion. Upon consensus, the payee can then utilize this confirmation as verifiable proof of the transaction 

\subsection{Peer to peer transaction}
\begin{figure}[htp]  
\centering
\includegraphics[width=0.9\columnwidth, height=1in]{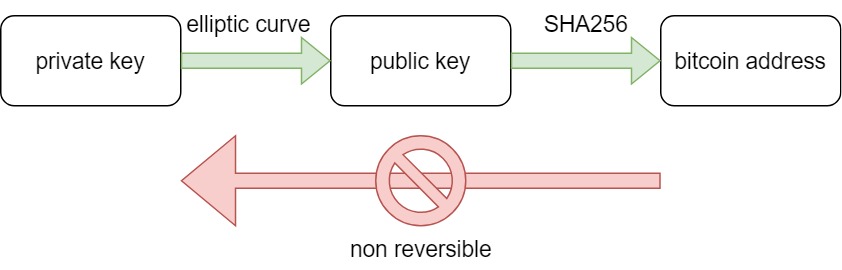}
\caption{key pair encryption}
\label{fig:keypair_encryption}
\end{figure}
Prior to executing a Bitcoin transaction, a user generates a cryptographic key pair comprising a private key and a public key. The private key consists of random digits and remains exclusively accessible to its owner. Conversely, the public key is derived from the private key through elliptic curve multiplication \cite{bos2014elliptic} and is publicly viewable across the network. When initiating a transaction, the payer specifies a Bitcoin address, serving as the recipient's designation, which is derived from the public key via cryptographic hashing using SHA256. Both elliptic curve multiplication and SHA256 hashing are unidirectional encryption processes, as illustrated in Figure ~\ref{fig:keypair_encryption}; the sole method to decrypt the private key entails exhaustive brute-force enumeration \cite{courtois2014optimizing}. Within this encryption framework, a digital signature produced by the private key can be authenticated by the public key without compromising the confidentiality of the private key. Subsequent to transaction creation, the sender generates a hashed signature and appends the recipient's hashed public key. This transaction is then broadcast to the nearest Bitcoin network node, which subsequently relays it to multiple proximate nodes. Through exponential growth, all Bitcoin nodes typically receive this transaction within a few seconds. This dissemination process is termed 'broadcasting,' and a transaction can only be appended to the ledger if it is approved by nearly all nodes. Ultimately, the node engages in the mining process, consolidating a batch of transactions into a singular block, subsequently appending this newly mined block to the existing blockchain. Thereafter, the transactions are rendered transparent to all nodes within the network, precluding any further modifications.
\subsection{Unspent Transaction Output Model}

\begin{figure}[htp]  
\centering
\includegraphics[width=0.9\columnwidth, height=1in]{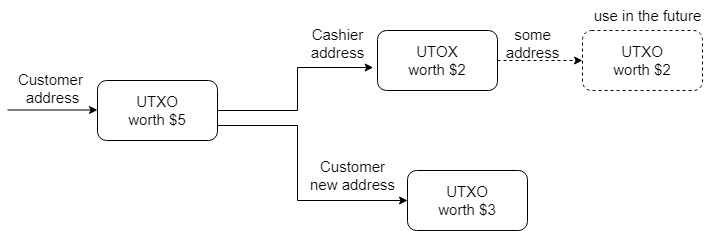}
\caption{Transactions are made with UTXO and every movement is recorded in the ledger}
\label{fig:utxo}
\end{figure}

In contrast to conventional fiat currencies, Bitcoin lacks denominational units and does not operate through accounts or balances within its ledger. Instead, the Unspent Transaction Output Model (UTXO) \cite{delgado2019analysis} is employed to represent indivisible currency units that are scatterd throughout the Bitcoin network. Each UTXO is associated with a specific address, and a wallet determines the balance by scanning the network and aggregating the distinct UTXO components. The conceptualization of UTXO can be analogized to purchasing goods at a retail store using cash and receiving change in return. For instance, when buying a \$2 soda at a store with a \$5 bill, the cashier would return \$3 in one-dollar bills as change. In the realm of Bitcoin, a similar transaction involving a \$5 UTXO as input would result in the creation of two outputs: one comprising a \$2 UTXO directed to the cashier's address, and another consisting of a \$3 UTXO sent to an address associated with the payer. As illustrated in Figure ~\ref{fig:utxo}, each transaction is meticulously recorded, detailing the UTXO amounts and the corresponding input and output addresses. These transactions are sequentially appended to the blockchain, constituting the immutable ledger of the Bitcoin network.

\subsection{Proof of work}
Bitcoin's goal, fully decentralized exchange, is primarily facilitated through the mining process. In the absence of a central authority, consensus mechanisms are crucial to safeguard against fraudulent transactions. The Proof of Work (PoW) algorithm \cite{fullmer2018analysis} serves as the foundational mechanism for achieving this consensus. Miners, which are servers operating Bitcoin network nodes, are responsible for validating transactions and appending them to the Bitcoin ledger. By aggregating a sufficient number of transactions into a block and subsequently appending this block to the blockchain, miners are incentivized through two distinct mechanisms: firstly, through the generation of new bitcoins via the coinbase transaction, and secondly, through transaction fees associated with the block in question.

\begin{figure}[htp]  
\centering
\includegraphics[width=0.9\columnwidth]{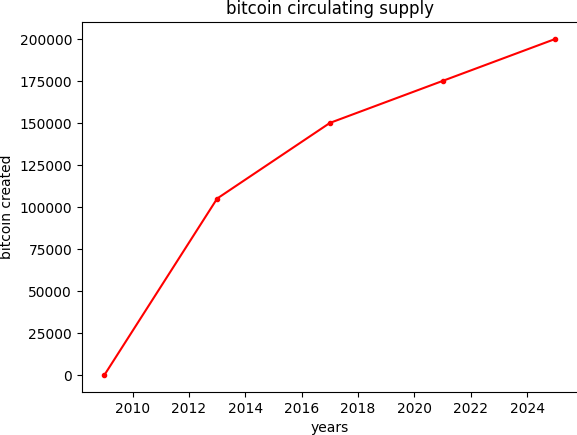}
\caption{New Bitcoin generated rate decreases by half every four year}
\label{fig:btc_supply}
\end{figure}

The mining process is characterized by the generation of new bitcoins, and the computational effort expended to solve cryptographic puzzles is termed as the 'Proof of Work'. Given the predetermined and finite supply of 21 million bitcoins, mining rewards undergo a halving process \cite{meynkhard2019fair} approximately every 210,000 blocks, resulting in a reduction of 50\%. Given that each block is generated, on average, every 10 minutes, this supply adjustment occurs approximately every four years. For instance, Starting from 2009, when the first block was mined, the mining incentives were 50 bitcoin. By April 2024, this reward had decreased to 3 bitcoins. As depicted in Figure ~\ref{fig:btc_supply}, which illustrates the circulating supply of Bitcoin over time, it is projected that nearly all bitcoins will be mined by the year 2140.

The algorithm behind the mining is hash calculation with an arbitrary data as input and form a fixed length of hash output that meets the target. The target\cite{o2014bitcoin} is the threshold for the output and only the outputs smaller than it can be used as the answer. In short words, the target asks the miner to find a hash with starting with number of zero bits. It is nearly impossible for one to guess the correct answer and the only way is to repeatedly hash the input with an incrementing parameter called nonce. It is noted that with SHA256 a fixed input and a fixed nonce will always output the same hash so that every other miner can verify if this output meets the target with one calculation. For example, we have an arbitrary input as "mine the bitcoin" and we want to find an output that less than the target 
\begin{lstlisting}[breaklines,breakindent=0pt,language=encryption]
0x1000000000000000000000000000000000000000000000000000000000000000
\end{lstlisting}
we start the nonce at 1 and increment by one every time before hashing. the input will be "mine the bitcoin \{nonce\}"
\begin{lstlisting}[breaklines,breakindent=0pt, language=encryption]
nonce: 0 hash: 46ba4680f2b00e708e6c6c043e191c261e14798218b0267e546b9285e9c05bb3
nonce: 1 hash: a7d726c49f8d462bfaf46d95bca63de90d4d7f0db90fb3979fc4d8cade027197
nonce: 2 hash: 5b558fd78b6a19697ba4364ad82fa2e05a4d22b416830db3db239160c4fa3557
nonce: 3 hash: d7cc60ade12ecc618b6fb8a4e6012a783e5c83791f1c15459ac1f96663ef5975
nonce: 4 hash: 0567f52f7dfc0a8f14a74c5825715e937d574e2140fe8889d25f61a9de0fb0a4
nonce: 5 hash: f3633b07da67961f6c55ec0d57e4758d4c90be972d135c4805422c3cae465506
nonce: 6 hash: 05b472655c3427d8dbbd6f812cda3048868b3e43102dab5ed006b5420f1ad419
\end{lstlisting}
when nonce equals 4, we get the first answer that meets the target because it starts with one zero. The mining difficulty is relatively low when the target is set as above. However, if we change the target that requires the output hash starts with more zeros, the mining difficulty can grows exponentially because there is no short-cut only solution to apply is bump the nonce for every calculation.

\section{Ethereum - Next generation blockchain platform}
Launched in 2015, Ethereum currently ranks as the second-largest cryptocurrency by market capitalization \cite{ranganthan2018decentralized}. While both Bitcoin and Ethereum are founded upon blockchain technology, their respective focuses and functionalities diverge significantly. Bitcoin primarily emphasizes decentralized exchange capabilities, whereas Ethereum enables more functionality, including smart contracts and decentralized applications. A core objective of Ethereum is to transcend the limitations inherent in Bitcoin's scripting language \cite{o2017simplicity} by enhancing it to achieve full Turing-completeness. This enhancement facilitates the execution of complex computational tasks, including loop. As a blockchain platform, Ethereum affords its users a more expansive range of capabilities beyond merely facilitating decentralized exchanges.

\subsection{Ethereum accounts}
\begin{figure}[htp]  
\centering
\includegraphics[width=1\linewidth, height=2.5in]{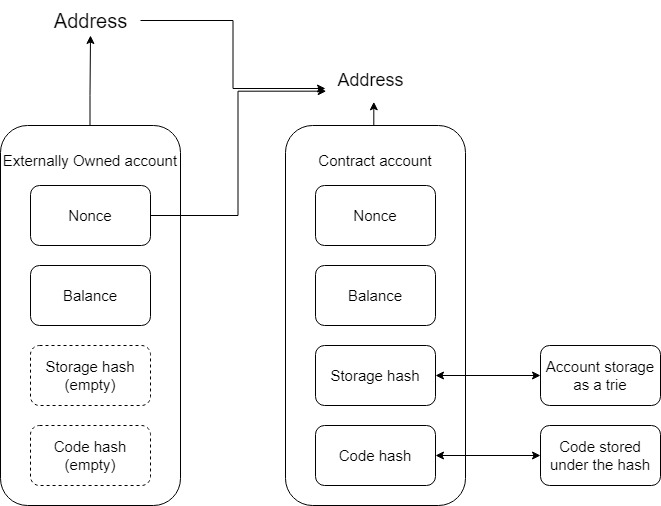}
\caption{Externally owned account and Contract account have the same structure}
\label{fig:ethaccount}
\end{figure}

Within the Ethereum ecosystem, state transitions are intrinsically linked to accounts. The Ethereum network recognizes two distinct types of accounts \cite{wood2014ethereum}, each distinguished by a unique 20-byte address. The first category, Externally Owned Accounts (EOA), operates akin to Bitcoin's model, functioning based on a key pair paradigm. Conversely, Contract Accounts operate based on their contract code.As illustrated in Figure ~\ref{fig:ethaccount}, Ethereum accounts comprise four primary fields. The 'Nonce' serves as a counter, indicating either the number of transactions initiated by an Externally Owned Account or the count of contracts created by a Contract Account. The 'Balance' field denotes the quantity of Ethereum currency (ETH) held within the account. The 'Storage Hash' is a 256-bit hash representing the trie root node, utilized for storing the account's content. Lastly, the 'Code Hash' refers to a hash value associated with the code segments programmed within the account.
\subsection{Transaction}
Unlike Bitcoin's UTXO model, Ethereum uses Ether(ETH) as a currency unit. ETH can be sent and collected during transactions and is also a form of payment for transaction fee. In Ethereum, there are three types of transaction. 
\begin{itemize}
    \item Regular transactions between each account
    \item Contract deployment transactions happens when a contract is deployed without a forwading address
    \item Execution of a contract is a transaction to invoke the program in a deployed contract address.
\end{itemize}
When a transaction is executed within the Ethereum network, analogous to Bitcoin, it is broadcast to the public and subsequently incorporated into a block by network nodes. A fee, denominated in ETH, is calculated and levied upon the sender. However, the methodology for fee calculation in Ethereum diverges from that of Bitcoin. In Ethereum, this fee is termed 'gas' and serves as a metric indicating the computational effort requisite for executing specific operations within the network. The gas fee can be conceptualized through a mathematical equation: \textbf{amount of effort expended} multiplied by \textbf{gas price per unit}. Furthermore, the gas fee is inherently variable, fluctuating in response to changes in the unit price, which is dynamically adjusted based on network demand \cite{donmez2022transaction}. Consequently, during periods of heightened popularity and increased network congestion, the gas fee tends to escalate, and conversely, it decreases during less congested periods. The implementation of the gas fee system enhances the security of the Ethereum network by deterring malicious activities that might exploit the network and by preventing the occurrence of infinite loops that could potentially deplete network resources.

\subsection{Smart contract}

It is noted that Ethereum contracts are not business-oriented entities; rather, they function as sets of rules that are executed upon invocation by a message or transaction. Smart contracts possess a balance and interact with transactions, constituting a distinct type of account within the Ethereum network. These smart contracts autonomously execute predefined rules through embedded programming \cite{metcalfe2020ethereum}. Importantly, once deployed, smart contracts are immutable and neither the transactions they execute nor the contract itself can be reversed.

For instance, in a real-world scenario, a landlord could deploy a smart contract to facilitate a rental system. Within this smart contract, the landlord could set a fixed monthly rent and a specified due date. Subsequently, the smart contract would automatically execute each month to collect the predetermined rent from the tenant. Failure on the part of the tenant to remit payment, or submission of an incorrect amount, could trigger eviction proceedings. Furthermore, smart contracts can also establish an escrow mechanism for tenant deposits, ensuring that these funds remain inaccessible until the termination of the lease agreement.

However, smart contracts are not without limitations, primarily stemming from their incapacity to interact with off-chain data sources. Returning to the previous rental example, rental rates may fluctuate in response to real-world variables. The inherent design of smart contracts precludes them from autonomously accessing off-chain data to effectuate such adjustments. To circumvent this limitation, specialized tools known as oracles are employed to facilitate the integration of external data sources with the Ethereum blockchain.

\subsection{Proof of stake}
Ethereum has pioneered a novel consensus mechanism known as Proof of Stake (PoS). Distinct from the previously discussed Proof of Work (PoW) system, PoS removes the need for solving computational puzzles. Within the Ethereum network, nodes have the opportunity to participate as validators, selected at random, upon staking a minimum of 32 ETH within a designated smart contract to authenticate transactions. Validators are incentivized through transaction fees and penalized from destroying their stakes if they engage in fraudulent activities.

\begin{figure}[htp]  
\centering
\includegraphics[width=1\linewidth, height=2.5in]{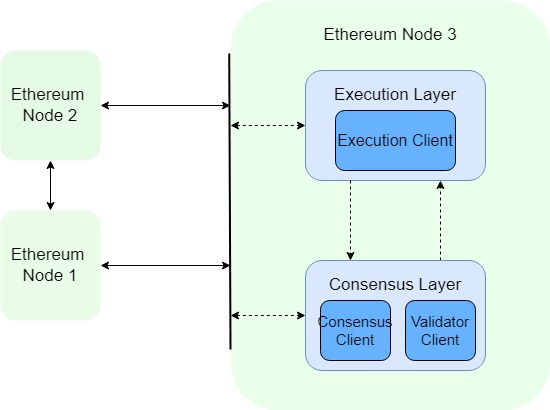}
\caption{Nodes have two layers as their clients communicates through different layers}
\label{fig:ethLayer}
\end{figure}

Upon the creation and signing of a transaction by a user, a gas fee is computed, serving as compensation for the selected validator responsible for generating and appending a block to the network \cite{nguyen2019proof}. As illustrated in Figure ~\ref{fig:ethLayer}, the transaction undergoes scrutiny by an Ethereum execution client to ascertain the validity of the signature and to confirm the sufficiency of funds. Subsequently, the execution client disseminates the transaction to the public via the execution layer, whereupon the randomly selected validator works on adding such transaction to a block.

Within the validator's node, the execution client aggregates multiple transactions, executing them as a batch to verify state changes, and subsequently conveys this information to the consensus client. Other nodes receive the block via consensus layer and then use their execution client to re-run those transaction locally for verification. If no mistakes are found, the validator client will play it role to attest the block is valid and this block is ready to be added to Ethereum blockchain.

This transition from PoW to PoS enhances both cost-effectiveness and energy efficiency, as PoS obviates the need for the continuous operation of high-performance CPUs tasked with solving computational puzzles, a hallmark feature of the PoW mechanism.

\section{Solana - High performance blockchain}
 
With the proliferation of decentralized applications and the concomitant surge in transaction volumes, the imperative for high-performance blockchain platforms has become increasingly salient. Solana, launched in 2020, was conceptualized as an innovative architecture tailored to meet the demands of high-throughput blockchain operations. A distinctive feature of Solana's design is the utilization of timestamps to validate decentralized transactions, a mechanism that markedly increases block creation throughput.

\begin{figure}[htp]
\centering

\includegraphics[width=0.9\columnwidth]{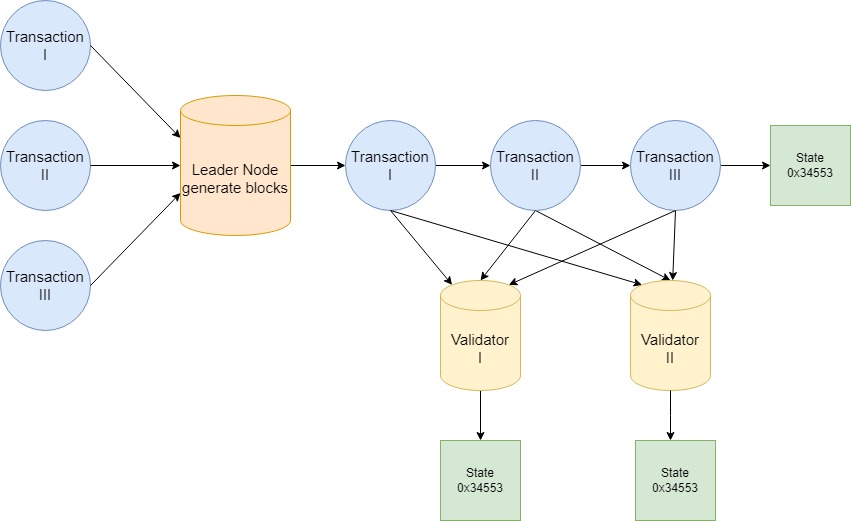}
\caption{Proof of History Sequence}
\label{fig:solanaTxn}
\end{figure}

Yakovenko \cite{yakovenko2018solana} explores that preceding blockchain implementations eschewed reliance on timestamps, thereby failing to ensure consistent decision-making across network nodes. Solana introduces the 'Proof of History' protocol, designed to construct a globally synchronized ledger with a verifiable passage of time. As illustrated in Figure ~\ref{fig:solanaTxn}, a node is selected at random to serve as the leader, entrusted with the task of block generation. The leader node sequentially processes transactions, generating a hash based on preceding hashes with each iteration. Subsequently, other nodes within the network can verify if the output hash is correct  by recalculating the hash value from the transaction I as long as the system chooses a non-collision hash method.
\subsection{Proof of history}
In contrast to Proof of work in bitcoin that takes about 10 minutes to generate the block, Proof of history algorithm solves this long waiting period by simultaneously creating and verifying the state in time ordering. In this case, time ordering is used as truth of source so the validators can still working on verification even they receive incorrect ordering. Table \ref{tab:Solana_hash} provides an overview of hash sequence. We firstly take the random string "Poh transaction input data" as an input for hashing; then we take the hashed value as input for the next hash operation. The leader node does this operation perpetually and appends latest transaction as input \(sha256(append(df986bf947..., sequence 4))\). Since the sha256 is a non-collision hash method, the output hash values are always unique and consistent in time ordering. This ensures that leader node is not able to guess or make up a fake output hash overhead. On the other hand, leader node can not fasten the process by utilizing multi-thread because of lack knowledge of previous output hash.

\begin{table}[!tbp] 
\centering
{
\vspace{0.20in}
\caption{
overview of proof of history sequence hash
}

\label{tab:Solana_hash}
 \resizebox{\columnwidth}{!}{
\begin{tabular}{ccccccc}
    \toprule
    \multicolumn{1}{c}{ Sequence Id} & \multicolumn{1}{c}{ Operation} & \multicolumn{1}{c}{ Output Hash} \cr
    
    \midrule
    1 & sha256("Poh transaction input data") & 5186766972...\cr
    2 & sha256("5186766972...") & 9e881c985e...\cr
    3 & sha256("9e881c985e...") & df986bf947...\cr
    \bottomrule
    \end{tabular}
}
}
\end{table}

\begin{figure}[htp]  
\centering
\includegraphics[width=0.7\columnwidth]{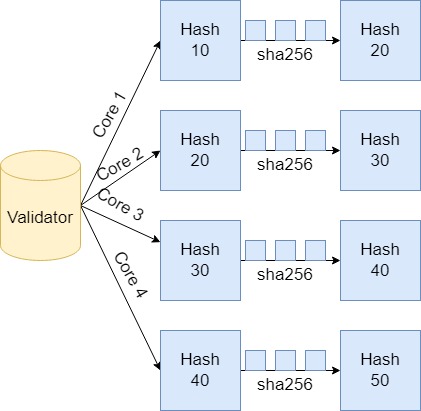}
\caption{output hashes are verified by multi-threading in significantly less time}
\label{fig:poh_validation}
\end{figure}

When it comes to the validation part, a number of nodes selected as validators is responsible for verifying the final output hash. The verificaiton is straight forward as recalculating every hash from the start and check if the final output hash is same as the one generated from the leader node. It is noted that validation is able to be accelerated by enabling multi-threading. Although the output hash is based on previous hash and is unpredictable, every hash has been calculated and stored at this time and validators can re-calculate it in parallel. As depicted in \ref{fig:poh_validation}, the validator utilizes four cores to verify four segments of output hashes. If Hash 20 output calculated is different than Hash 20 input, the validator will reject the leader's work and rest hash calculations follows the same pattern. As Yakovenko\cite{yakovenko2018solana} indicates that these verifications are preferred to be implemented on GPU and modern GPU contains 4000 cores. Such run in parallel verifications are expected to only consume 1/4000 time compare to the generation time

\subsection{Consensus mechanism}
As previously stated, Solana has incorporated the Proof of History mechanism for state transition and further integrates the Proof of Stake consensus mechanism for validations. Each validator is required to stake a certain amount of funds as collateral during the validation process. Analogous to Ethereum, Solana imposes both penalties and rewards to foster a constructive voting system. A pivotal concept introduced within this consensus mechanism is that of a 'Super Majority,' representing \(\dfrac{2}{3}\)  of validators weighted by their staked funds. Should a node endeavor to subvert the network by submitting invalid votes, the attacker would necessitate control over more than \(\dfrac{1}{3}\) funds on the network. For instance, consider a scenario where three validators, denoted as V1, V2, and V3, each stake funds worth 100 dollars. Assuming V1 acts maliciously, the overall validation remains legitimate due to the super majority attained by V2 and V3, who collectively stake 200 dollars out of a total of 300 dollars. Conversely, if V1 were to stake 200 dollars, the validation would become invalid, as V2 and V3 would lack the requisite weighted funds. To mitigate such scenarios, the network initiates a timeout period during which the three validators engage in repeated verification attempts. Should V1 consistently fail to validate during these attempts, it will be expelled from the validator group, and its staked funds will be confiscated.

\section{Conclusion}
 Bitcoin, Ethereum, and Solana stand as prominent cryptocurrencies in the market. Each of these platforms introduces distinctive technological innovations designed to address and overcome the inherent limitations of preceding blockchain systems. Bitcoin pioneered the Proof of Work consensus mechanism, which eradicates the dependence on third-party financial intermediaries but inadvertently engenders the consumption of substantial electricity resources through the intensive computation puzzles. Ethereum, on the other hand, introduced the Proof of Stake algorithm to facilitate an economically efficient validation process. Nonetheless, the platform continues to struggle with block creation speeds compares to centralized transaction processors. In contrast, Solana's groundbreaking hybrid integration of the Proof of Stake and Proof of History mechanisms has significantly enhanced transaction throughput capabilities, offering a promising solution to the scalability challenges encountered by earlier blockchain architectures.




%



\newpage
\bibliographystyle{IEEEtran}
\bibliography{references}

\end{document}